\documentclass{amsart}
\usepackage{amssymb}
\usepackage{amsfonts}
\usepackage{graphicx}
\usepackage{epstopdf}
\usepackage[backref]{hyperref}

\newcommand{\beq}{\begin{equation}}
\newcommand{\eeq}{\end{equation}}

\begin{document}

\title{On the map of Vogel's Plane}
 
 \author{R. L. Mkrtchyan}
       \address{Yerevan Physics Institute, 2 Alikhanian Brothers St., Yerevan, 0036, Armenia}
           \email{mrl55@list.ru}      
  
\maketitle




{\small  {\bf Abstract.} We search points in a Vogel plane with regular universal expression for character of adjoint representation. This gives seven patterns of singularities cancellation, each giving a certain Diophantine equation of third order on three variables. Solutions of these equations are classical series of simple Lie algebras (including an ``odd symplectic'' one), $D_{2,1,\lambda}$ superalgebra, the straight line of three-dimensional algebras, and a number of isolated solutions, including exceptional simple Lie algebras. One of these Diophantine equations, namely knm=4k+4n+2m+12 contains all simple Lie algebras, except SO(2N+1). Isolated solutions  contain, beside exceptional simple Lie algebras, so called $E_{7\frac{1}{2}}$ algebra and also two other  similar (unknown) objects with positive dimensions. In addition, there are 47 isolated solutions in ``unphysical semiplane''  with negative dimensions. Isolated solutions mainly lie on a few straight lines in Vogel plane. All solutions give an integers in  universal dimension formulae for first three symmetric powers of adjoint representation.}

\section{Introduction}

In papers \cite{V0,V} Vogel introduced a so-called Universal Lie algebra, with the motivations  from knot theory and aimed ``to construct a monoidal category which looks like
the category of module over a Lie algebra and which is universal in some sense`` \cite{V}.  It is not clear  what are the final conclusions (if any) of that approach from the point of view of initial aims, but it appears that some new and very interesting formulae arose, both in Lie algebras and in physical theories \cite{LM1,LM2,LM3,LM4,MSV,MV}. Another source of almost the same formulae is the Delign's et.al \cite{Del,DM} approach of series of Lie algebras. These formulae are  developed for exceptional Lie algebras and coincide with those of universal Lie algebra when latter are restricted on an exceptional algebras.  From physics side, one may think on a Universal Lie algebra as a  generalisation of  famous and exceptionally fruitful 't Hooft's \cite{H1} idea of analytic continuation of  SU(N) gauge theories  from integer to an arbitrary N (and subsequent 1/N expansion) to  exceptional Lie algebras. Let's consider one of the first universal expressions, namely that for the dimensions of simple Lie algebras \cite{V0}:

\beq
\label{f3}
dim \, \mathfrak {g} = \frac{(\alpha-2t)(\beta-2t)(\gamma-2t)}{\alpha\beta\gamma}.
\eeq
where parameters for all simple Lie algebras are given in a Table \ref{tab:par} below.

\begin{table}[ht]  
\caption{Vogel's parameters for simple Lie algebras}     
\begin{tabular}{|c|c|c|c|c|c|}
\hline
Type & Lie algebra  & $\alpha$ & $\beta$ & $\gamma$  & $t=\alpha+\beta+\gamma=h^\vee$\\   
\hline    
$A_n$ &  $\mathfrak {sl}_{n+1}$     & $-2$ & 2 & $(n+1) $ & $n+1$\\
$B_n$ &   $\mathfrak {so}_{2n+1}$    & $-2$ & 4& $2n-3 $ & $2n-1$\\
$C_n$ & $ \mathfrak {sp}_{2n}$    & $-2$ & 1 & $n+2 $ & $n+1$\\
$D_n$ &   $\mathfrak {so}_{2n}$    & $-2$ & 4 & $2n-4$ & $2n-2$\\
$G_2$ &  $\mathfrak {g}_{2}  $    & $-2$ & $10/3 $& $8/3$ & $4$ \\
$F_4$ & $\mathfrak {f}_{4}  $    & $-2$ & $ 5$& $ 6$ & $9$\\
$E_6$ &  $\mathfrak {e}_{6}  $    & $-2$ & $ 6$& $ 8$ & $12$\\
$E_7$ & $\mathfrak {e}_{7}  $    & $-2$ & $ 8$& $ 12$ & $18$ \\
$E_8$ & $\mathfrak {e}_{8}  $    & $-2$ & $ 12$& $20$ & $30$\\
\hline  
\end{tabular}
\label{tab:par}
\end{table}

Here we see the characteristic features of universal expressions: they are functions (analytic, rational, etc.) on projective parameters $\alpha,\beta, \gamma$ belonging to Vogel plane - real projective space factorized w.r.t. the all permutations of these three parameters, which (functions) on the points from Table \ref{tab:par} give an answers for corresponding simple Lie algebras. So, in comparison with 1/N expansion we now have two parameters, which at some discrete points on Vogel plane give all simple Lie algebras. Would one  prove that gauge-invariant quantities in gauge theories depend on gauge groups only through universal parameters, then one can develop an approximation scheme, generalising 1/N. This have been proved for some quantities in Chern-Simons theory (perturbative partition function, unknotted adjoint Wilson average, etc.) in \cite{MV} and conjectured for other theories.

Actually there are a lot of open questions this approach is rising. One of them is the question on a ``population'' of Vogel space. Table \ref{tab:par} lists the  points of simple Lie algebras, question is whether there are  other interesting points?
 
In this paper we suggest an approach to study the ``road map'' of Vogel plane, i.e. to find an interesting, by some definition, points on that plane, as well as roads, by which we understand straight lines between them. Definition mentioned is the requirement that these ``populated'' points have a regular, in a finite complex $x$ plane, universal expression \cite{MV}, (\ref{gene}), for characters of adjoint representation of simple Lie algebras, evaluated at point $x\rho$ ($\rho$ is a Weyl vector). I.e. that are the points, sharing with simple Lie algebras the feature of regularity of their adjoint characters.  This definition appears to be very restrictive, mainly selecting points of simple Lie algebras of Table \ref{tab:par}, but with interesting additions. 

It is tempting to present things in a following way: Universal Lie algebra's  can be considered as an approach to classification of Lie algebras, which removes at the initial stage the requirement of dimensions of being integer. This is done by considering an invariant quantities, namely (vacuum Feynman, in physics language) diagrams constructed from trivalent vertex, implied to be a structure constants of some Lie algebra. This leads to existence of three parameters and universal formulae, expressing many objects - dimensions, eigenvalues, characters - through these three parameters by formulae which are senseful for almost all values of parameters. To finish classification, some requirement of integer-validness should be introduced. That may be e.g. a requirement of dimensions to be an integers. For example, this requirement, putted on exceptional line Exc in Vogel's plane (see below), restrict parameters on a special points on that line, corresponding to five exceptional simple Lie algebras, plus one more solution, so called  $E_{7\frac{1}{2}}$ \cite{W,LM3}. This algebra have a features similar to simple Lie algebras, particularly it gives integers in all universal dimension formulae. From this point of view requirement of character to be non-singular seems to be some general form of possible integer-validness requirements, and indeed it leads both to identification of points, corresponding to all simple Lie algebras, and to discovery of new points, similar to $E_{7\frac{1}{2}}$. 

Interestingly, this approach reveals a (perhaps previously unknown) connection of classification of simple Lie algebras with certain Diophantine equations.

If all three projective parameters are non-zero, then by permutations and projective transformations one can achieve the positivity of two of them, but then the sign of remaining one cannot be changed. Since universal formula (\ref{f3}) requires parameters to be non-zero, one naturally consider two semiplanes in Vogel plane - one with negative sign of third parameter (usually we choose that to be $\alpha$), and other one with all parameters positive. All Lie algebras belong to the first semiplane, as seen from Table, which for that reason we shall call ``physical''. In that semiplane the dimension formula can give a positive number, while in the other, ``nonphysical'' semiplane dimensions (\ref{f3}) always are negative.  

The most important formula we shall use is the universal expression for character $\chi_{ad}(x\rho)$ of adjoint representation evaluated at point $x\rho$ ($\rho$ - Weyl vector), discovered in \cite{MV}:

\begin{eqnarray} \label{gene}
\chi_{ad}(x\rho)=r+\sum_{\mu\in R}e^{x(\mu,\rho)}\equiv f(x)\\ 
f(x)=\frac{\sinh(x\frac{\alpha-2t}{4})}{\sinh(\frac{x\alpha}{4})}\frac{\sinh(x\frac{\beta-2t}{4})}{\sinh(x\frac{\beta}{4})}\frac{\sinh(x\frac{\gamma-2t}{4})}{\sinh(x\frac{\gamma}{4})} \label{gene2}
\end{eqnarray}

For simple Lie algebras the initial expression (\ref{gene}) for $f(x)$ is non-singular in finite complex plane of $x$, so the final expression (\ref{gene2}) should have same feature. So, as mentioned above, we are seeking all points in Vogel plane where $f(x)$ is regular in finite $x$ plane. 

 Two cases can be separated according to zero or non-zero value of one of the numbers $2t-\alpha, 2t-\beta, 2t-\gamma$. If it is zero and all parameters are non-zero, then character is zero. We call this line on a Vogel plane a 0d line. The case when in addition some of parameters $\alpha, \beta, \gamma$ are zero is indefinite 0/0 and one should define and calculate a limit. The answers for dimension (\ref{f3}) and character (\ref{gene2}) appear to  depend on definition and  this is considered at the end of Section \ref{srmp}. We shall  call this 0/0 case. The other possibility is that  $2t-\alpha, 2t-\beta, 2t-\gamma$ all are non-zero, which together with non-singularity of character means that all parameters are non-zero. Then, each zero of sins in denominator at some value of $x$ should be canceled by zero of nominator, which means that for each value of  $\kappa = \alpha, \beta, \gamma$ at least one of the ratios $(2t-\alpha)/\kappa, (2t-\beta)/\kappa,(2t-\gamma)/\kappa$ should be integer. 

The complete matrix of these ratios
$$R_{\kappa,\sigma}=(2t-\kappa)/\sigma, 
$$

where $\kappa, \sigma = \alpha, \beta, \gamma $, can be easily calculated for all simple Lie algebras, examples for SU(n) and $G_2$ are given below.

\begin{table}[ht]
\caption{Matrix R for SU(N)}
	\centering
		\begin{tabular}{|rrr|} 
		-(N+1)&1-N&$-\frac{N}{2}$  \\ 
		N+1&N-1&$\frac{N}{2}$\\ 
		$2+\frac{2}{N}$&$2-\frac{2}{N}$&1\\
		\end{tabular}
	
	\label{tab:KForSU}
\end{table}

\begin{table}[ht]
\caption{Matrix R for  G2}
	\centering
		\begin{tabular}{|r r r|} 
	-5&$-\frac{7}{3}$&$-\frac{8}{3}$\\ 
	3&$\frac{7}{5}$&$\frac{8}{5}$\\ 
	$\frac{15}{3}$&$\frac{7}{4}$&2	\\ 	
		\end{tabular}
	
	\label{tab:KForG2}
\end{table}

One can see that abovementioned feature is revealed, i.e. for both algebras in each row we have an integer. The same happens for all other simple Lie algebras. Moreover, there are a different mechanisms of cancellation, in a sense that we can have an integers in a different places in a given row. These places can be changed by permutation of parameters, but it is not possible to put them on a desired places in the rows.  We shall call a ``pattern'' a set of places, one in the each row, of matrix $k_{\kappa,\sigma}$ where integers are present.  The same algebra can appear in a different patterns. For example exceptional algebra $G_2$ appear in one abovementioned pattern, but $E_7$ in two others, also. It is easy to establish, that there are at most 7 different, up to permutations, patterns of matrix $k$ to have integers in each row. 

The specified property is necessary, but not enough, in principle, for character to be regular. Extra care should be provided for the cases when some parameters become integer simultaneously, say $\alpha$ and $\beta$. If that happens in the pattern with integer $k_{\kappa,\alpha}$ and $k_{\kappa,\beta}$ (same $\kappa$) then it appears that denominator in $f(x)$ has a second-order zero, which cannot be canceled by the one first order zero in nominator. But  we didn't met such a phenomena. More exactly, below we find  solutions for all patterns and explicitly check that $f(x)$ is regular, i.e. in all cases some other nominator ``automatically'' becomes integer. Some explanation of this phenomena is given in Section \ref{sdf}. 

So, we shall act as follows: assume that some pattern is realized, i.e. in each row of matrix $k_{\kappa,\sigma}$  there is at least one integer and try to obtain for what values of parameters it is possible. This requirement turns to be very restrictive, forcing parameters to be fixed almost exactly to the Table \ref{tab:par} above. But, interestingly, there are some exceptions. Some of them were known earlier ($E_{7\frac{1}{2}}$ and ``symplectic algebras with half-integer rank'' \cite{Pr}),  others are new. 

The abovementioned seven patterns are the following, all others can be obtained by permutations, neither two from these seven are connected by permutation. To list them we list three values of $\kappa$, i.e. $(\kappa_1,\kappa_2,\kappa_3)$ for which $k_{\kappa_1,\alpha},k_{\kappa_2,\beta},k_{\kappa_3,\gamma}$  are integers (denote them $k,n,m$ respectively):

$$1)(\alpha,\alpha,\alpha), 2)(\alpha,\alpha,\beta), 3)(\alpha,\alpha,\gamma),4)(\alpha,\beta,\gamma), 5)(\alpha,\gamma,\beta),6)(\beta,\alpha,\alpha),7)(\beta,\gamma,\alpha)$$
We shall call them 1aaa, 2aab, 3aag, 4abg, 5agb, 6baa and 7bga respectively. 

Consider for example the fourth, most symmetric, pattern 4abg. One have 

\beq \label{4p}
(2t-\alpha)=k\alpha, (2t-\beta)=n\beta,(2t-\gamma)=m\gamma
\eeq

or in matrix form in Table \ref{tab:MatrixForm}.

\begin{table}[h]
\caption{Matrix form of Eq.(\ref{4p})}
	\centering
		\begin{tabular}{|rrr||r|r} 
		1-k &2 &2&$ \alpha $  \\ 
		2& 1-n & 2 & $\beta$ & =0 \\ 
		2&2&1-m& $\gamma $ \\
			\end{tabular}
	\label{tab:MatrixForm}
\end{table}

This is a system of three linear equations on three variables $\alpha,\beta,\gamma$, with zero variable-free terms, so non-trivial solution can exist only if corresponding determinant is zero. That determinant  is a third order polynomial over k,n,m and equation is $knm=kn+nm+km+3n+3k+3m+5$. We shall call this equation a Diophantine equation (or condition) for a given pattern. Below it is presented in another form, as a condition that the sum of three rational fraction is equal to  unity (provided that corresponding denominators are non-zero), which is another form of identity $\alpha+\beta+\gamma=t$. In this form this Diophantine condition looks classically, although we didn't find discussion of exactly this equation. Another, normalized form, i.e. that without second order terms can be achieved by variables shift $k\rightarrow k+1,n\rightarrow n+1, m\rightarrow m+1$, which leads to $knm=4m+4n+4k+16$. Below are presented a Diophantine condition in first two forms  and expressions for universal parameters through k,n,m:

\begin{eqnarray} 
\label{abg}
knm=kn+nm+km+3n+3k+3m+5 \\  
\label{abg2}
\frac{2}{k+1}+\frac{2}{n+1}+\frac{2}{m+1}=1 \\ 
\label{abg3}
\alpha=\frac{2t}{k+1}, \beta=\frac{2t}{n+1}, \gamma=\frac{2t}{m+1}  
\end{eqnarray}

The Diophantine condition (\ref{abg}) can be solved in a different ways. One can note that modules of all three parameters can't be simultaneously larger than 5. Indeed, then l.h.s. of (\ref{abg2}) is less than 6/7. There is a solution with k=l=m=5, actually this will be a solution for all 7 patterns.  Then let's separate the case of singular denominator(s), i.e. when one of integers say m is -1: then from (\ref{abg}) we have (k+1)(n+1)=0, i.e. then one other integer also should be -1, and remaining one is arbitrary. This is the case when two rows or columns  of the matrix of Eq. (\ref{4p}) (Table \ref{tab:MatrixForm}) are proportional, so determinant is zero independently from the value of remaining parameter in third row or column. In other form this case can be described as follows: let's represent Eq. (\ref{abg}) as say $mx+y=0$, where $x, y$ are second order over $k,n$, and seek solutions for system $x=y=0$. In the case under consideration it happens at $k=n=-1$. Then (\ref{abg2}) becomes an identity, i.e. it is satisfied at any m. Finally, when all three integers k, n, m are -1 the third row (column) is also proportional to remaining ones, so the rank of matrix is one. In that case two parameters from three $\alpha,\beta,\gamma$ remain arbitrary. 

Next assume that none of integers is -1. Since  equations (\ref{abg}) is symmetric w.r.t. the permutations of k,n,m, which leads to corresponding permutation of $\alpha,\beta,\gamma$ in (\ref{abg3}), we can restrict $|k|\geq |n|\geq |m|$, so we have $|m|\leq 5$. 
Then from (\ref{abg2})
$$\frac{2}{k+1}+\frac{2}{n+1}=1-\frac{2}{m+1}
$$
The minimal value of modulus of r.h.s. of this eq. is zero, at m=1, and next minimal is 1/3, at m=2. If it is zero, then m=1 and k=-n-2, n arbitrary. If it is not the case, then assuming that both k and n are larger in modulus than some N, we have 
$$\frac{4}{N+1}\geq |\frac{2}{k+1}+\frac{2}{n+1}|=|1-\frac{2}{m+1}| \geq \frac{1}{3}
$$
from which we get $N\leq 11$ hence $n \leq 11$ (remember our convention $k \geq n $). Finally, for remaining case with none of k, n, m equal $\pm 1$ we have 
 
$$\frac{2}{k+1}=1-\frac{2}{m+1}-\frac{2}{n+1}
$$
 
 The minimal value of modulus of r.h.s. is 1/21 at n=6, m=2, so the maximal positive value of k is 41, minimal negative -43. This gives a solution (41,6,2). Now it is not difficult to find all solutions, either by continuation of similar considerations, or, since the maximal values of modules of k, n, m are restricted, by going over all the cases with the help of computer program, which we actually did. 

So, altogether we come to the following conclusions for pattern 4abg, which actually will be similar for all patterns. Solutions of equation (\ref{abg}) are of three kinds. Two kinds are a series, which means that some of k,n,m can be arbitrarily large, third one consist from an isolated solutions. 

 The first kind of series (we shall call them classical ones)  appear when at some value of one of the integers equation (\ref{abg}) becomes linear, here it happens at say m=1:
$$
n+k+2=0
$$
and solution for (k,n,m) is (-n-2,n,1), with arbitrary integer n, giving $(\alpha,\beta,\gamma)=(-2t/(n+1),2t/(n+1),t)$ which corresponds to SU(n+1) algebra. At n=-1 we have k=-1 and  solution is $t=\gamma=\alpha+\beta=0$, so these solutions together can represented as if first one is multiplied on (n+1)/t:  $(\alpha,\beta,\gamma)=(-2,2,n+1)$. Similar phenomena will happen for other patterns, also.  The second kind series (call them non-classical)  appear when two from three integers, say k and n, become -1. Then (\ref{abg}) becomes an identity, i.e. it is satisfied at any m.  This solution at m=-1 gives a rank=1 matrix in (\ref{4p}) and correspondingly leads to an arbitrary parameters $(\alpha,\beta,\gamma)$ with restriction t=0  which represents $D_{2,1,\lambda}$ superalgebra. At $m\neq-1$ solution is t=0 and  $\gamma=0$, this is 0/0 case. These series solutions are listed in Table \ref{tab:series}.

Isolated solutions  for this pattern in physical semiplane consist from SO(8) with $(k,n,m)=(2,2,-7)$ and $(\alpha,\beta,\gamma)=(-2,-2,1)$, $E_6$ with (k,n,m)=(3,2,-13) and $(\alpha,\beta,\gamma)=(-3,-4,1)$, $E_8$ with $(k,n,m)=(4,2,-31)$ and $(\alpha,\beta,\gamma)=(-6,-10,1)$, plus two points: $0d_3$ with $(k,n,m)=(0,	-5,	-5)$ and $(\alpha,\beta,\gamma)=(4,	-1,	-1)$, and $0d_4$ with $(k,n,m)=(0,	-4,	-7)$ and $(\alpha,\beta,\gamma)=(6,	-2,	-1)$, both with dimensions zero (hence notation, all $0d_i$ listed in Tables below). There is also a number of  points in unphysical semiplane. All isolated solutions are listed  in Table \ref{tab:4abg} below.

\begin{scriptsize}
\begin{table}[ht]
\caption{Points in Vogel plane: series}
	\centering
		\begin{tabular}{|r|r|r|r|r|r|}  \hline
		$\mathfrak{g}$&$\alpha,\beta,\gamma$&Patterns&k,n,m&dim&rank\\ \hline
		SU(N)&2,-2,N&3aag&N-1,-N+1,1 &$N^2-1$&N-1\\
				&-2,2,N&4abg&-N-1,N-1,1&&\\
				&N,-2,2&5agb&1,1-N,N+1&&\\
		SU(2N)&N,1,-1 &1aaa&1,-N,N&$(2N)^2-1$&2N-1\\
					&&2aab&1,N,1-2N&&\\
					&&3aag&1,N,-2N-1&&\\
	SO(2k+4)&2,k,-1&3aag&k,2,-2k-3&(k+2)(2k+3)&2k+2\\
					&2,-1,k&1aaa&k,-2k,2&&\\
					&2,k,-1&2aab&k,2,-2-k&&\\
	SO(2N+1)&-2,4,2N-3&2aab&-2N,N,2&N(2N+1)&N\\	
	&-4,2,3-2N&6baa&N,3-2N,2&&\\
	SO(4N) &-1,2,2N-2&5agb&1-4N,N,2&2N(4N-1)&2N\\
	&2N-2,2,-1&6baa&2,N,-2N&&\\
	&-1,2,2N-2&6baa&N,2,4-4N&&\\
	$D_{2,1,\lambda}$&$\alpha+\beta+\gamma=0$&4abg&-1,-1,-1&1&1\\
			&-n, 1, (n-1)&3aag&-1,n,-1&&\\
	3d & $2\alpha+\beta+\gamma=0$ &5agb & -3,1,1 &3&1\\
	   	&-n,3,2n-3&2aab&-3,n,1&&\\
	&-2,-2k,k+1&3aag&k,1,-3&&\\		
	&  -m-2,m,1&6baa&1,1,m&&\\ 
	0d & $\alpha+2\beta+2\gamma=0$&1aaa &0,0,0& 0 &0 \\ 
			&	2(m+1),-(m+3),2&3aag& 0,0,m&& \\   
			&2,2-2k,2k-3&6baa& k,0,0&& \\   
			& 2m-2,-m-2,3&2aab&0,0,m&&\\ \hline
					\end{tabular}
	\label{tab:series}
\end{table}
\end{scriptsize}

\begin{scriptsize}
\begin{table}[ht]
\caption{Isolated solutions of 4abg pattern}
	\centering
		\begin{tabular}{|r|r|r|r|r|r|} \hline
		k n m& $\alpha \beta \gamma$ & Dim & Rank & Algebra& Lines\\ \hline
4	2	-31	&	-6	-10	1	&	248	&	8	&	$	E_8	$	&	Exc, M	\\
3	2	-13	&	-3	-4	1	&	78	&	6	&	$	E_6	$	&	Exc, F	\\
2	2	-7	&	-2	-2	1	&	28	&	4	&	$	SO(8)	$	&	SO,Exc	\\
0	-5	-5	&	4	-1	-1	&	0	&	0	&	$	0d_3	$	&	0d	\\
0	-4	-7	&	6	-2	-1	&	0	&	0	&	$	0d_4	$	&	0d	\\
5	5	5	&	1	1	1	&	-125	&	-19	&	$	Y_1	$	&	?	\\
9	4	4	&	1	2	2	&	-144	&	-14	&	$	Y_{10}	$	&	K,M	\\
7	7	3	&	1	1	2	&	-147	&	-17	&	$	Y_{11}	$	&	F	\\
11	5	3	&	1	2	3	&	-165	&	-13	&	$	Y_{15}	$	&	F	\\
19	4	3	&	1	4	5	&	-228	&	-10	&	$	Y_{29}	$	&	F,M	\\
11	11	2	&	1	1	4	&	-242	&	-18	&	$	Y_{31}	$	&	Exc	\\
14	9	2	&	2	3	10	&	-252	&	-8	&	$	Y_{35}	$	&	Exc	\\
17	8	2	&	1	2	6	&	-272	&	-14	&	$	Y_{38}	$	&	Exc	\\
23	7	2	&	1	3	8	&	-322	&	-12	&	$	Y_{43}	$	&	Exc	\\
41	6	2	&	1	6	14	&	-492	&	-10	&	$	Y_{47}	$	&	Exc	\\
\hline
	\end{tabular}
	
	\label{tab:4abg}
\end{table}
\end{scriptsize}

Solutions for other patterns are presented in a next Section in a  similar  form. We list  Diophantine equations, their series solutions and isolated solutions, and identify an algebras when it is possible. The cases when denominators are zero are taken into account separately, when it is needed. The Diophantine conditions for all patterns - both in initial and normalised (i.e. without second order terms) forms are given in Table \ref{tab:Diophantine}. Series solution for all patterns are combined in Table \ref{tab:series}. Isolated solutions are given in corresponding Tables for each pattern. In additional columns of these Tables we present a dimensions, according to (\ref{f3}), rank, i.e. a constant term in a finite expansion (\ref{gene}), and lines, to which that point belongs to (see Section \ref{srmr}). Notations for isolated solutions are as follows - those in a physical  semiplane  are denoted as $X_i$, in unphysical as $Y_i$, $i=1,2,...$ in the decrease of dimensions order. When dimensions coincide, but universal parameters doesn't, we of course use different notations. In principle such objects can coincide,   despite the different sets of parameters, as SU(2) - (-2,2,2)  and SO(3) - (-2,4,-1). Perhaps all points on a 3d curve should be considered as corresponding to the algebras isomorphic to SU(2), since characters are $f(x) = (q^2+1+q^{-2})$, $q=e^{x\gamma/4}$ . But in all cases with coinciding dimensions larger (in modulus) than three but different points in Vogel plane  we check that their characters are different, so they are really different objects. In most cases already ranks are different, as can be easily seen from corresponding tables. Below we always use notation SO(N) with both positive and negative N, in the latter case even N give an Sp(N) algebras \cite{Ki,Mkr,Cvitbook,VM}, so we never use notation Sp below.

\section{Road map: population.} \label{srmp}

For the first pattern 1aaa we have 
\beq \label{1p}
(2t-\alpha)=k\alpha, (2t-\alpha)=n\beta,(2t-\alpha)=m\gamma
\eeq

Zero determinant condition and solution (provided denominators are non-zero):

\begin{eqnarray}  \label{aaa}
knm=2kn+nm+2km \\ 
\label{aaa2}
\frac{2}{k+1}+\frac{2k}{n(k+1)}+\frac{2k}{m(k+1)}=1 \\ 
\label{aaa3}
\alpha=\frac{2t}{k+1}, \beta=\frac{2tk}{n(k+1)}, \gamma=\frac{2tk}{m(k+1)}  
\end{eqnarray}

These conditions are symmetric w.r.t. the transposition of n and m, leading to the transposition of $\beta,\gamma$. 
The classical series can appear at k=1, then m=-n, and finally $(\alpha,\beta,\gamma)=(n,1,-1)$ (also taking into account n=0 possibility as mentioned in 4abg pattern above), which corresponds to SU(2n). Other possibility  for classical series is m=2 (or n=2). Then n=-2k,  and similarly $(\alpha,\beta,\gamma)=(2,-1,k)$ which corresponds to SO(2k+4). Nonclassical series appear when two from three integers k,n,m are zero. Then, if third integer is zero also, the rank of matrix is one and we have a 0d line $2t-\alpha=0$, otherwise it is the 0/0 case and is discussed at the end of this Section. All essentially different (i.e. without transpositions of n and m) isolated solutions are listed  in Table \ref{tab:1aaa} below.

\begin{scriptsize}

\begin{table}[ht]
\caption{Isolated solutions of 1aaa pattern}
	\centering
		\begin{tabular}{|r|r|r|r|r|r|} \hline
		k n m& $\alpha \beta \gamma$ & Dim & Rank & Algebra& Lines\\ \hline
2	3	-12	&	-6	-4	1	&	133	&	7	&	$	E_7	$	&	Exc, T	\\
-15	5	3	&	1	-3	-5	&	99	&	7	&	$	X_2	$	&	T	\\
-6	4	3	&	2	-3	-4	&	21	&	3	&	$	SO(7)	$	&	SO,T,K	\\
-3	1	-3	&	-1	3	-1	&	3	&	1	&	$	3d_1	$	&	3d	\\
-3	3	3	&	1	-1	-1	&	3	&	1	&	$	SU(2)	$	&	3d,T	\\
-2	1	-4	&	-2	4	-1	&	3	&	1	&	$	SO(3)	$	&	3d	\\
5	5	5	&	1	1	1	&	-125	&	-19	&	$	Y_1	$	&	?	\\
6	6	4	&	2	2	3	&	-132	&	-10	&	$	Y_4	$	&	K	\\
4	8	4	&	2	1	2	&	-144	&	-14	&	$	Y_{10}	$	&	K,M	\\
3	6	6	&	2	1	1	&	-147	&	-17	&	$	Y_{11}	$	&	F	\\
10	5	4	&	2	4	5	&	-153	&	-7	&	$	Y_{13}	$	&	K	\\
9	9	3	&	1	1	3	&	-189	&	-17	&	$	Y_{21}	$	&	T	\\
3	12	4	&	4	1	3	&	-195	&	-11	&	$	Y_{23}	$	&	F,K	\\
6	12	3	&	2	1	4	&	-195	&	-13	&	$	Y_{24}	$	&	T	\\
12	8	3	&	2	3	8	&	-207	&	-7	&	$	Y_{26}	$	&	T	\\
5	15	3	&	3	1	5	&	-221	&	-11	&	$	Y_{28}	$	&	T	\\
2	8	8	&	4	1	1	&	-242	&	-18	&	$	Y_{31}	$	&	Exc	\\
2	12	6	&	6	1	2	&	-272	&	-14	&	$	Y_{38}	$	&	Exc	\\
21	7	3	&	1	3	7	&	-285	&	-11	&	$	Y_{39}	$	&	T	\\
4	24	3	&	6	1	8	&	-319	&	-9	&	$	Y_{42}	$	&	T,M	\\
2	20	5	&	10	1	4	&	-377	&	-11	&	$	Y_{45}	$	&	Exc	\\
 \hline

	\end{tabular}
	
	\label{tab:1aaa}
\end{table}
\end{scriptsize}

For the second pattern 2aab we have 
\beq \label{2p}
(2t-\alpha)=k\alpha, (2t-\alpha)=n\beta,(2t-\beta)=m\gamma
\eeq

Zero determinant condition and solution (provided denominators are non-zero):

\begin{eqnarray} \label{aab}
knm=2kn+nm+2km+2n-2k \\ 
\label{aab2}
\frac{2}{k+1}+\frac{2k}{n(k+1)}+\frac{2(nk+n-k)}{mn(k+1)}=1 \\ 
 \label{aab3}
\alpha=\frac{2t}{k+1}, \beta=\frac{2tk}{n(k+1)}, \gamma=\frac{2t(nk+n-k)}{mn(k+1)}  
\end{eqnarray}

The series appear at k=1, m=1-2n, giving $(\alpha,\beta,\gamma)=(n,1,-1)$ which is SU(2n), then at m=2, k=-2n, $(\alpha,\beta,\gamma)=(2t/(1-2n),4t/(2n-1),t(3-2n)/(1-2n)\propto (-2,4,2n-3)$ which is SO(2n+1), and at n=2, m=-2-k $(\alpha,\beta,\gamma)=(2t/(1+k),tk/(1+k),-t/(1+k) \propto (2,k,-1)$ which is SO(2k+4). Nonclassical series appear at k=n=0 with $(\alpha,\beta,\gamma)\propto(2m-2,-m-2,3)$, which belongs to 0d line; or n=0, m=1 with $(\alpha,\beta,\gamma)\propto(0,-\gamma,\gamma)$ which belongs both to 0d line and to case 0/0; or k=0, m=-2 with $(\alpha,\beta,\gamma)\propto(-2\gamma,0,\gamma)$ which belongs both to 0d line and to case 0/0; or k=-3,m=1. This last case corresponds to the points with $3\alpha+n\beta=0$ (arbitrary integer $n$) on the  line $2\alpha+\beta+\gamma=0$ with dim=3. Actually all  points on the line $2\alpha+\beta+\gamma=0$ correspond to non-singular function f(x) with dim=3. That can be checked directly: 

\begin{eqnarray} 
f(x)=\frac{\sinh(x\frac{\alpha-2t}{4})}{\sinh(\frac{x\alpha}{4})}\frac{\sinh(x\frac{\beta-2t}{4})}{\sinh(x\frac{\beta}{4})}\frac{\sinh(x\frac{\gamma-2t}{4})}{\sinh(x\frac{\gamma}{4})} \\
=\frac{\sinh(x\frac{3\alpha}{4})}{\sinh(\frac{x\alpha}{4})}\frac{\sinh(x\frac{-\gamma}{4})}{\sinh(x\frac{\beta}{4})}\frac{\sinh(x\frac{-\beta}{4})}{\sinh(x\frac{\gamma}{4})}\\
=e^{\frac{x\alpha}{2}}+1+e^{-\frac{x\alpha}{2}}
\end{eqnarray}

Isolated solutions are listed in the Table.

\begin{scriptsize}
\begin{table}[ht]
\caption{Isolated solutions of 2aab pattern}
	\centering
		\begin{tabular}{|r|r|r|r|r|r|} \hline
		k n m& $\alpha \beta \gamma$ & Dim & Rank & Algebra& Lines\\ \hline
2	-16	5	&	-8	1	-5	&	190	&	8	&	$	E_{7\frac{1}{2}}	$	&	Exc	\\
4	-16	3	&	-4	1	-7	&	156	&	8	&	$	X_1	$	&	T, M	\\
2	3	-14	&	-6	-4	1	&	133	&	7	&	$	E_7	$	&	Exc, T	\\
-15	3	3	&	1	-5	-3	&	99	&	7	&	$	X_2	$	&	T	\\
2	-6	4	&	-6	2	-5	&	52	&	4	&	$	F_4	$	&	Exc, K	\\
3	-6	3	&	-2	1	-3	&	45	&	5	&	$	SO(10)	$	&	SO,T,F	\\
5	5	5	&	1	1	1	&	-125	&	-19	&	$	Y_1	$	&	?	\\
4	5	6	&	10	8	7	&	-129	&	-1	&	$	Y_2	$	&	M	\\
6	6	4	&	2	2	3	&	-132	&	-10	&	$	Y_4	$	&	K	\\
7	4	5	&	4	7	5	&	-135	&	-3	&	$	Y_7	$	&	K	\\
4	4	8	&	2	2	1	&	-144	&	-14	&	$	Y_{10}	$	&	K,M	\\
3	6	7	&	2	1	1	&	-147	&	-17	&	$	Y_{11}	$	&	F	\\
12	4	4	&	2	6	5	&	-168	&	-6	&	$	Y_{16}	$	&	K	\\
4	12	4	&	6	2	7	&	-184	&	-6	&	$	Y_{17}	$	&	K,M	\\
10	8	3	&	4	5	13	&	-186	&	-2	&	$	Y_{18}	$	&	T	\\
9	3	7	&	1	3	1	&	-189	&	-17	&	$	Y_{21}(1)	$	&	T	\\
9	9	3	&	1	1	3	&	-189	&	-17	&	$	Y_{21}(2)	$	&	T	\\
3	4	13	&	4	3	1	&	-195	&	-11	&	$	Y_{23}(2)	$	&	F,K	\\
3	12	5	&	4	1	3	&	-195	&	-11	&	$	Y_{23}(1)	$	&	F,K	\\
6	3	10	&	2	4	1	&	-195	&	-13	&	$	Y_{24}	$	&	T	\\
12	3	6	&	2	8	3	&	-207	&	-7	&	$	Y_{26}	$	&	T	\\
15	6	3	&	2	5	9	&	-207	&	-5	&	$	Y_{27}	$	&	T	\\
5	3	13	&	3	5	1	&	-221	&	-11	&	$	Y_{28}	$	&	T	\\
7	14	3	&	2	1	5	&	-231	&	-13	&	$	Y_{30}	$	&	T	\\
2	8	11	&	4	1	1	&	-242	&	-18	&	$	Y_{31}	$	&	Exc	\\
2	9	10	&	18	4	5	&	-245	&	-3	&	$	Y_{33}	$	&	Exc	\\
2	6	16	&	6	2	1	&	-272	&	-14	&	$	Y_{38}	$	&	Exc	\\
21	3	5	&	1	7	3	&	-285	&	-11	&	$	Y_{39}	$	&	T	\\
25	5	3	&	1	5	7	&	-285	&	-9	&	$	Y_{40}	$	&	T	\\
2	14	8	&	14	2	5	&	-296	&	-6	&	$	Y_{41}	$	&	Exc	\\
4	3	22	&	6	8	1	&	-319	&	-9	&	$	Y_{42}	$	&	T,M	\\
6	24	3	&	4	1	9	&	-342	&	-10	&	$	Y_{44}	$	&	T	\\
2	5	26	&	10	4	1	&	-377	&	-11	&	$	Y_{45}	$	&	Exc	\\
2	24	7	&	12	1	5	&	-434	&	-10	&	$	Y_{46}	$	&	Exc	\\
 \hline
	\end{tabular}
	\label{tab:2aab}
\end{table}
\end{scriptsize}

For the third pattern we have

\beq \label{3p}
(2t-\alpha)=k\alpha, (2t-\alpha)=n\beta,(2t-\gamma)=m\gamma
\eeq

Zero determinant condition and solution (provided denominators are non-zero):

\begin{eqnarray} 
 \label{aag}
kmn=mn+2km+kn+3n+2k\\  
\label{aag2}
\frac{2}{k+1}+\frac{2k}{n(k+1)}+\frac{2}{m+1}=1 \\
\alpha=\frac{2t}{k+1}, \beta=\frac{2tk}{n(k+1)}, \gamma=\frac{2t}{m+1}  \label{aag3}
\end{eqnarray}

The classical series appear at k=1, m=-2n-1, $(\alpha,\beta,\gamma)=(t,t/n,-t/n)$, handling n=0 case separately, we get altogether $(\alpha,\beta,\gamma)=(n,1,-1)$,  which is SU(2n); at m=1, n=-k, $(\alpha,\beta,\gamma)=(2t/(1+k),-2t/(1+k),t) \propto (2,-2,k+1)$ which is SU(k+1), and at n=2, m=-2k-3, $(\alpha,\beta,\gamma)=(2t/(k+1),tk/(k+1),-t/(k+1)) \propto (2,k,-1)$ which is SO(2k+4). The non-classical series are (k,m)=(0,-3), (n,m)=(0,-1), (k,n)=(-1,1), all correspond to  0/0 case, (k,n)=(0,0) for which  $(\alpha,\beta,\gamma)=(2(m+1),-(m+3),2)$ which belong to 0d line. Two other non-classical series are:  (k,m)=(-1,-1) giving points (-n, 1, (n-1)) on the line t=0, i.e.  $D_{2,1,\lambda}$ superalgebra, and finally  (n,m)=(1,-3), which gives $\beta-k\alpha=0$ points on the 3d line $\alpha+\beta+2\gamma=0$.
Isolated solutions are listed in the Table \ref{tab:3aag}. 

\begin{scriptsize}
\begin{table}[ht]
\caption{Isolated solutions of 3aag pattern}
	\centering
		\begin{tabular}{|r|r|r|r|r|r|} \hline
		k n m& $\alpha \beta \gamma$ & Dim & Rank & Algebra& Lines\\ \hline
2	-20	4	&	-10	1	-6	&	248	&	8	&	$	E_8(1)	$	&	Exc, M	\\
4	-24	2	&	-6	1	-10	&	248	&	8	&	$	E_8(2)	$	&	Exc, M	\\
-25	5	2	&	1	-5	-8	&	190	&	8	&	$	E_{7\frac{1}{2}}	$	&	Exc	\\
-21	3	4	&	1	-7	-4	&	156	&	8	&	$	X_1	$	&	T, M	\\
2	3	-19	&	-6	-4	1	&	133	&	7	&	$	E_7	$	&	Exc, T	\\
2	-8	3	&	-4	1	-3	&	78	&	6	&	$	E_6(1)	$	&	Exc, F	\\
3	-9	2	&	-3	1	-4	&	78	&	6	&	$	E_6(2)	$	&	Exc, F	\\
-10	4	2	&	2	-5	-6	&	52	&	4	&	$	F_4	$	&	Exc, K	\\
-9	3	3	&	1	-3	-2	&	45	&	5	&	$	SO(10)	$	&	SO,T,F	\\
2	-4	2	&	-2	1	-2	&	28	&	4	&	$	SO(8)	$	&	SO,Exc	\\
-5	3	2	&	3	-5	-4	&	14	&	2	&	$	G_2	$	&	Exc, T	\\
-9	-3	0	&	-1	-3	8	&	0	&	0	&	$	0d_1	$	&	0d	\\
-6	-4	0	&	-2	-3	10	&	0	&	0	&	$	0d_2	$	&	0d	\\
-5	-5	0	&	-1	-1	4	&	0	&	0	&	$	0d_3	$	&	0d	\\
-4	-8	0	&	-2	-1	6	&	0	&	0	&	$	0d_4	$	&	0d	\\
-2	4	0	&	2	-1	-2	&	0	&	0	&	$	0d_5	$	&	K,0d	\\
3	-1	0	&	-1	3	-4	&	0	&	0	&	$	0d_6	$	&	F,0d	\\
5	5	5	&	1	1	1	&	-125	&	-19	&	$	Y_1	$	&	?	\\
4	6	5	&	6	4	5	&	-130	&	-4	&	$	Y_3	$	&	M	\\
6	4	6	&	2	3	2	&	-132	&	-10	&	$	Y_4	$	&	K	\\
7	5	4	&	5	7	8	&	-132	&	-2	&	$	Y_5	$	&	M	\\
5	4	7	&	4	5	3	&	-133	&	-2	&	$	Y_6	$	&	K	\\
8	4	5	&	2	4	3	&	-140	&	-8	&	$	Y_9	$	&	K	\\
4	4	9	&	2	2	1	&	-144	&	-14	&	$	Y_{10}(1)	$	&	K,M	\\
4	8	4	&	2	1	2	&	-144	&	-14	&	$	Y_{10}(2)	$	&	K,M	\\
3	6	7	&	2	1	1	&	-147	&	-17	&	$	Y_{11}(1)	$	&	F	\\
7	7	3	&	1	1	2	&	-147	&	-17	&	$	Y_{11}(2)	$	&	F	\\
3	7	6	&	7	3	4	&	-150	&	-4	&	$	Y_{12}(1)	$	&	F	\\
6	8	3	&	4	3	7	&	-150	&	-4	&	$	Y_{12}(2)	$	&	F	\\
3	5	9	&	5	3	2	&	-153	&	-7	&	$	Y_{14}(1)	$	&	F	\\
9	6	3	&	2	3	5	&	-153	&	-7	&	$	Y_{14}(2)	$	&	F	\\
3	9	5	&	3	1	2	&	-165	&	-13	&	$	Y_{15}(1)	$	&	F	\\
5	10	3	&	2	1	3	&	-165	&	-13	&	$	Y_{15}(2)	$	&	F	\\
14	4	4	&	2	7	6	&	-184	&	-6	&	$	Y_{17}	$	&	K,M	\\
7	3	11	&	3	7	2	&	-187	&	-7	&	$	Y_{20}	$	&	T	\\
9	3	9	&	1	3	1	&	-189	&	-17	&	$	Y_{21}	$	&	T	\\
3	4	15	&	4	3	1	&	-195	&	-11	&	$	Y_{23}(1)	$	&	F,K	\\
6	3	13	&	2	4	1	&	-195	&	-13	&	$	Y_{24}	$	&	T	\\
15	5	3	&	1	3	4	&	-195	&	-11	&	$	Y_{23}(2)	$	&	F,K	\\
11	3	8	&	3	11	4	&	-200	&	-4	&	$	Y_{25}	$	&	T	\\
5	3	17	&	3	5	1	&	-221	&	-11	&	$	Y_{28}	$	&	T	\\
3	15	4	&	5	1	4	&	-228	&	-10	&	$	Y_{29}(1)	$	&	F,M	\\
4	16	3	&	4	1	5	&	-228	&	-10	&	$	Y_{29}(2)	$	&	F,M	\\
15	3	7	&	1	5	2	&	-231	&	-13	&	$	Y_{30}	$	&	T	\\
2	8	11	&	4	1	1	&	-242	&	-18	&	$	Y_{31}(1)	$	&	Exc	\\
11	11	2	&	1	1	4	&	-242	&	-18	&	$	Y_{31}(2)	$	&	Exc	\\
10	12	2	&	6	5	22	&	-244	&	-2	&	$	Y_{32}	$	&	Exc	\\
2	7	13	&	14	4	3	&	-247	&	-5	&	$	Y_{34}	$	&	Exc	\\
2	10	9	&	10	2	3	&	-252	&	-8	&	$	Y_{35}	$	&	Exc	\\
15	9	2	&	3	5	16	&	-258	&	-4	&	$	Y_{37}	$	&	Exc	\\
2	6	17	&	6	2	1	&	-272	&	-14	&	$	Y_{38}(1)	$	&	Exc	\\
2	12	8	&	6	1	2	&	-272	&	-14	&	$	Y_{38}(2)	$	&	Exc	\\
8	16	2	&	2	1	6	&	-272	&	-14	&	$	Y_{38}(3)	$	&	Exc	\\
20	8	2	&	2	5	14	&	-296	&	-6	&	$	Y_{41}	$	&	Exc	\\
4	3	29	&	6	8	1	&	-319	&	-9	&	$	Y_{42}	$	&	T,M	\\
2	16	7	&	8	1	3	&	-322	&	-12	&	$	Y_{43}(1)	$	&	Exc	\\
7	21	2	&	3	1	8	&	-322	&	-12	&	$	Y_{43}(2)	$	&	Exc	\\
27	3	6	&	1	9	4	&	-342	&	-10	&	$	Y_{44}	$	&	T	\\
2	5	29	&	10	4	1	&	-377	&	-11	&	$	Y_{45}	$	&	Exc	\\
35	7	2	&	1	5	12	&	-434	&	-10	&	$	Y_{46}	$	&	Exc	\\
2	28	6	&	14	1	6	&	-492	&	-10	&	$	Y_{47}(1)	$	&	Exc	\\
6	36	2	&	6	1	14	&	-492	&	-10	&	$	Y_{47}(2)	$	&	Exc	\\
\hline
	\end{tabular}
		\label{tab:3aag}
\end{table}
\end{scriptsize}

Note that $E_6$ and $E_8$ appear twice, as well as some $Y_i$. We denote them e.g. $E_6(1), E_6(2)$, and similarly in other cases when some algebra appears two or three times in the same pattern. These are different descriptions of the same algebra. That these descriptions are really different, can be understood e.g. from the expression for dimension of algebra  through k, n, m. These formulae are given in Section \ref{sdf}, for 3aag pattern it is 
$$dim(g)= m(k-n-kn)  $$
Thus, for e.g. $E_8$ the same dimension's integer 248 is represented as  $4\times 62$ for $E_8(1)$ and as  $2\times 124$ for $E_8(2)$. Moreover, as mentioned in Conclusion the $k \leftrightarrow n$ symmetry of the normalized form of Diophantine condition (\ref{aag}) interchanges $E_8(1)$ and $E_8(2)$ with different algebras.

For the fifth pattern we have 
\beq \label{5p}
(2t-\alpha)=k\alpha, (2t-\gamma)=n\beta,(2t-\beta)=m\gamma
\eeq

Zero determinant condition and solution (provided denominators are non-zero):

\begin{eqnarray} \label{agb}
knm=-5 - 3 k + 2 m + 2 k m + 2 n + 2 k n + m n \\
\label{agb2}
\frac{2}{k+1}+\frac{2(m-1)}{mn-1}+\frac{2(n-1)}{mn-1}=1 \\ 
\label{agb3}
\alpha=\frac{2t}{k+1}, \beta=\frac{2t(m-1)}{mn-1}, \gamma=\frac{2t(n-1)}{mn-1}  
\end{eqnarray}

These conditions are symmetric w.r.t. the transposition of n and m, leading to the transposition of $\alpha,\beta$. 
The classical series appear at k=1,n=1-N,m=N+1, $(\alpha,\beta,\gamma)=(t,-2t/N,2t/N) \propto (N,-2,2)$, which is SU(N). Another series solution is at m=2 (or n=2). Taking  m=2, k=1-4N,n=N, we get $(\alpha,\beta,\gamma)=(-t/(2N-1),2t/(2N-1),2t(N-1)/(2N-1)) \propto (-1,2,2N-2)$, which is SO(4N).

Non-classical series are at (n,m)=(1,1) and  (k,n)=(-3,1) (or (k,m)=(-3,1))  which coincide at point (k,n,m)=(-3,1,1) and give a 3d line $2\alpha+\beta+\gamma=0$, otherwise they give a 0/0 case. 

Isolated solutions are listed in the Table \ref{tab:5agb}.

\begin{scriptsize}
\begin{table}[ht]
\caption{Isolated solutions of 5agb pattern}
	\centering
		\begin{tabular}{|r|r|r|r|r|r|} \hline
		k n m& $\alpha \beta \gamma$ & Dim & Rank & Algebra& Lines\\ \hline
2	5	-19	&	-8	-5	1	&	190	&	8	&	$	E_{7\frac{1}{2}}	$	&	Exc	\\
4	3	-13	&	-4	-7	1	&	156	&	8	&	$	X_1	$	&	T, M	\\
3	3	-5	&	-2	-3	1	&	45	&	5	&	$	SO(10)	$	&	SO,T,F	\\
-1	-1	-1	&	2	-1	-1	&	1	&	1	&	$	SO(2)	$	&	SO,$D_{2,1,\lambda}$	\\
0	-5	-5	&	4	-1	-1	&	0	&	0	&	$	0d_3	$	&	0d	\\
0	-3	-11	&	8	-3	-1	&	0	&	0	&	$	0d_1	$	&	0d	\\
0	7	-1	&	-4	-1	3	&	0	&	0	&	$	0d_6	$	&	F,0d	\\
5	5	5	&	1	1	1	&	-125	&	-19	&	$	Y_1	$	&	?	\\
5	7	4	&	3	2	4	&	-140	&	-8	&	$	Y_9	$	&	K	\\
9	4	4	&	1	2	2	&	-144	&	-14	&	$	Y_{10}	$	&	K,M	\\
3	7	7	&	2	1	1	&	-147	&	-17	&	$	Y_{11}	$	&	F	\\
9	7	3	&	1	1	3	&	-189	&	-17	&	$	Y_{21}	$	&	T	\\
3	13	5	&	4	1	3	&	-195	&	-11	&	$	Y_{23}	$	&	F,K	\\
13	5	3	&	1	2	4	&	-195	&	-13	&	$	Y_{24}	$	&	T	\\
7	11	3	&	2	1	5	&	-231	&	-13	&	$	Y_{30}	$	&	T	\\
2	11	11	&	4	1	1	&	-242	&	-18	&	$	Y_{31}	$	&	Exc	\\
21	4	3	&	1	4	6	&	-252	&	-10	&	$	Y_{36}	$	&	T,K	\\
6	19	3	&	4	1	9	&	-342	&	-10	&	$	Y_{44}	$	&	T	\\
2	31	7	&	12	1	5	&	-434	&	-10	&	$	Y_{46}	$	&	Exc	\\
 \hline
	\end{tabular}
	
	\label{tab:5agb}
\end{table}
\end{scriptsize}

For the sixth pattern we have 
\beq \label{6p}
(2t-\beta)=k\alpha, (2t-\alpha)=n\beta,(2t-\alpha)=m\gamma
\eeq

Zero determinant condition and solution (provided denominators are non-zero):

\begin{eqnarray} \label{baa}
knm=2kn+2nm+2km-2n-3m \\
\label{baa2}
\frac{2(n-1)}{kn-1}+\frac{2(k-1)}{kn-1}+\frac{2(kn-n)}{m(kn-1)}=1 \\ 
\label{baa3}
\alpha=t\frac{2(n-1)}{kn-1}, \beta=t\frac{2(k-1)}{kn-1}, \gamma=t\frac{2(kn-n)}{m(kn-1)}  
\end{eqnarray}

The series can appear at k=2, m=-2n, $(\alpha,\beta,\gamma)=(2t(n-1)/(2n-1),2t/(2n-1),-t/(2n-1) \propto (2n-2,2,-1))$ which is SO(4n), or at n=2, m=4-4k, $(\alpha,\beta,\gamma)=(-1,2,2k-2)$  which is SO(4k), or at m=2, n=3-2k, $(\alpha,\beta,\gamma)=(-4,2,3-2k)$ which is SO(2k+1).

Non-classical series appear at (n,m)=(0,0), giving (2,2-2k,2k-3) points on 0d line,  (k,m)=(1,0) and (n,m)=(1,-2), both belonging to 0/0 case , (k,n)=(1,1) which gives an (-m-2,m,1) points on the 3d line $2\gamma+\alpha+\beta=0$. 

Isolated solutions are listed in the Table \ref{tab:6baa}.

\begin{scriptsize}
\begin{table}[ht]
\caption{Isolated solutions of 6baa pattern}
	\centering
		\begin{tabular}{|r|r|r|r|r|r|} \hline
		k n m& $\alpha \beta \gamma$ & Dim & Rank & Algebra& Lines\\ \hline
-11	5	3	&	-5	1	-3	&	99	&	7	&	$	X_2(1)	$	&	T	\\
-9	3	5	&	-3	1	-5	&	99	&	7	&	$	X_2(2)	$	&	T	\\
3	-9	3	&	-3	-5	1	&	99	&	7	&	$	X_2(3)	$	&	T	\\
-3	3	4	&	-3	2	-4	&	21	&	3	&	$	SO(7)	$	&	SO,T,K	\\
-3	-3	1	&	3	-1	-1	&	3	&	1	&	$	3d_1	$	&	3d	\\
-1	-5	1	&	5	-3	-1	&	3	&	1	&	$	3d_2	$	&	3d,T	\\
-1	3	3	&	-1	1	-1	&	3	&	1	&	$	SU(2)(1)	$	&	3d,T	\\
3	-1	1	&	-1	-1	1	&	3	&	1	&	$	SU(2)(2)	$	&	3d,T	\\
5	5	5	&	1	1	1	&	-125	&	-19	&	$	Y_1	$	&	?	\\
4	5	6	&	5	8	6	&	-132	&	-2	&	$	Y_6	$	&	K	\\
6	6	4	&	3	2	2	&	-132	&	-10	&	$	Y_4	$	&	K	\\
5	7	4	&	7	6	4	&	-135	&	-3	&	$	Y_8	$	&	K	\\
4	4	8	&	1	2	2	&	-144	&	-14	&	$	Y_{10}	$	&	K,M	\\
9	5	4	&	5	2	4	&	-153	&	-7	&	$	Y_{13}	$	&	K	\\
4	10	4	&	5	6	2	&	-168	&	-6	&	$	Y_{16}	$	&	K	\\
3	6	8	&	3	10	4	&	-186	&	-4	&	$	Y_{19}(1)	$	&	T	\\
6	3	10	&	3	4	10	&	-186	&	-4	&	$	Y_{19}(2)	$	&	T	\\
3	7	7	&	1	3	1	&	-189	&	-17	&	$	Y_{21}(1)	$	&	T	\\
7	3	9	&	1	1	3	&	-189	&	-17	&	$	Y_{21}(2)	$	&	T	\\
7	11	3	&	11	5	3	&	-189	&	-3	&	$	Y_{22}	$	&	T	\\
9	9	3	&	3	1	1	&	-189	&	-17	&	$	Y_{21}(3)	$	&	T	\\
3	5	10	&	1	4	2	&	-195	&	-13	&	$	Y_{24}(1)	$	&	T	\\
5	3	12	&	1	2	4	&	-195	&	-13	&	$	Y_{24}(2)	$	&	T	\\
3	9	6	&	3	8	2	&	-207	&	-7	&	$	Y_{26}(1)	$	&	T	\\
9	3	8	&	3	2	8	&	-207	&	-7	&	$	Y_{26}(2)	$	&	T	\\
3	4	16	&	1	6	4	&	-252	&	-10	&	$	Y_{36}(1)	$	&	T,K	\\
4	3	18	&	1	4	6	&	-252	&	-10	&	$	Y_{36}(2)	$	&	T,K	\\
3	15	5	&	3	7	1	&	-285	&	-11	&	$	Y_{39}(1)	$	&	T	\\
5	21	3	&	7	5	1	&	-285	&	-9	&	$	Y_{40}	$	&	T	\\
15	3	7	&	3	1	7	&	-285	&	-11	&	$	Y_{39}(2)	$	&	T	\\
19	7	3	&	7	1	3	&	-285	&	-11	&	$	Y_{39}(3)	$	&	T	\\
 \hline
	\end{tabular}
	
	\label{tab:6baa}
\end{table}
\end{scriptsize}

For the seventh pattern we have 
\beq \label{7p}
(2t-\beta)=k\alpha, (2t-\gamma)=n\beta,(2t-\alpha)=m\gamma
\eeq

Zero determinant condition and solution (provided denominators are non-zero):

\begin{eqnarray} \label{bga}
knm=5 - 2 k - 2 m + 2 k m - 2 n + 2 k n + 2 m n \\ 
\label{bga2}
\frac{2(1 - m + m n)}{1 + k m n}+\frac{2 (1 - k + k m)}{1 + k m n}+\frac{2 (1 - n + k n)}{1 + k m n}=1 \\ 
\label{bga3}
\alpha=2t\frac{1 - m + m n}{1 + k m n}, \beta=2t\frac{1 - k + k m}{1 + k m n}, \gamma=2t\frac{1 - n + k n}{1 + k m n}  
\end{eqnarray}
This is cyclic symmetric: under $k \rightarrow n \rightarrow m \rightarrow k$ one have $ \alpha \rightarrow \beta \rightarrow \gamma \rightarrow \alpha $, correspondingly one can restrict really different solutions by, say, $k \geq n$ together with $k \geq m$, or in the case when two largest numbers coincide by say $k=n\geq m$.
There is no classical series for this pattern because corresponding linear equation when say k=2: $1+2m+2n=0$ has no integer solutions. Non-classical series would appear from the equation k(nm-2n-2m-2)=2mn-2n+5  when both coefficient at k and r.h.s should be put equal to zero which is not possible since r.h.s is odd integer. Isolated solutions are listed in the Table \ref{tab:7bga}.

\begin{scriptsize}
\begin{table}[ht]
\caption{Isolated solutions of 7bga pattern}
	\centering
		\begin{tabular}{|r|r|r|r|r|r|} \hline
		k n m& $\alpha \beta \gamma$ & Dim & Rank & Algebra& Lines\\ \hline
3	3	-11	&	-3	-5	1	&	99	&	7	&	$	X_2	$	&	T	\\
1	1	-3	&	1	-3	1	&	3	&	1	&	$	3d_1	$	&	3d	\\
3	-1	1	&	-1	1	-1	&	3	&	1	&	$	SU(2)	$	&	3d	\\
3	1	-1	&	1	-5	3	&	3	&	1	&	$	3d_2	$	&	3d,T	\\
5	5	5	&	1	1	1	&	-125	&	-19	&	$	Y_1	$	&		\\
9	3	7	&	3	11	5	&	-189	&	-3	&	$	Y_{22}	$	&	T	\\
9	7	3	&	1	1	3	&	-189	&	-17	&	$	Y_{21}	$	&	T	\\
19	3	5	&	1	7	5	&	-285	&	-9	&	$	Y_{40}	$	&	T	\\
19	5	3	&	1	3	7	&	-285	&	-11	&	$	Y_{39}	$	&	T	\\
 \hline
	\end{tabular}
		\label{tab:7bga}
\end{table}
\end{scriptsize}

We combine information about isolated points solutions in the physical Vogel semiplane in Table \ref{tab:is-phys}.

\begin{scriptsize}
\begin{table}[ht]
\caption{Isolated solutions in physical semiplane}
	\centering
		\begin{tabular}{|r|r|r|r|r|r|r|} \hline
		k n m& $\alpha \beta \gamma$ & Dim & Rank & Algebra& Lines&Patterns\\ \hline
2	-20	4	&	-10	1	-6	&	248	&	8	&	$	E_8(1)	$	&	Exc, M	&	3aag	\\
4	-24	2	&	-6	1	-10	&	248	&	8	&	$	E_8(2)	$	&	Exc, M	&	3aag	\\
4	2	-31	&	-6	-10	1	&	248	&	8	&	$	E_8	$	&	Exc, M	&	4abg	\\
2	-16	5	&	-8	1	-5	&	190	&	8	&	$	E_{7\frac{1}{2}}	$	&	Exc	&	2aab	\\
-25	5	2	&	1	-5	-8	&	190	&	8	&	$	E_{7\frac{1}{2}}	$	&	Exc	&	3aag	\\
2	5	-19	&	-8	-5	1	&	190	&	8	&	$	E_{7\frac{1}{2}}	$	&	Exc	&	5agb	\\
4	-16	3	&	-4	1	-7	&	156	&	8	&	$	X_1	$	&	T, M	&	2aab	\\
-21	3	4	&	1	-7	-4	&	156	&	8	&	$	X_1	$	&	T, M	&	3aag	\\
4	3	-13	&	-4	-7	1	&	156	&	8	&	$	X_1	$	&	T, M	&	5agb	\\
2	3	-12	&	-6	-4	1	&	133	&	7	&	$	E_7	$	&	Exc, T	&	1aaa	\\
2	3	-14	&	-6	-4	1	&	133	&	7	&	$	E_7	$	&	Exc, T	&	2aab	\\
2	3	-19	&	-6	-4	1	&	133	&	7	&	$	E_7	$	&	Exc, T	&	3aag	\\
-15	5	3	&	1	-3	-5	&	99	&	7	&	$	X_2	$	&	T	&	1aaa	\\
-15	3	3	&	1	-5	-3	&	99	&	7	&	$	X_2	$	&	T	&	2aab	\\
-11	5	3	&	-5	1	-3	&	99	&	7	&	$	X_2(1)	$	&	T	&	6baa	\\
-9	3	5	&	-3	1	-5	&	99	&	7	&	$	X_2(2)	$	&	T	&	6baa	\\
3	-9	3	&	-3	-5	1	&	99	&	7	&	$	X_2(3)	$	&	T	&	6baa	\\
3	3	-11	&	-3	-5	1	&	99	&	7	&	$	X_2	$	&	T	&	7bga	\\
2	-8	3	&	-4	1	-3	&	78	&	6	&	$	E_6(1)	$	&	Exc, F	&	3aag	\\
3	-9	2	&	-3	1	-4	&	78	&	6	&	$	E_6(2)	$	&	Exc, F	&	3aag	\\
3	2	-13	&	-3	-4	1	&	78	&	6	&	$	E_6	$	&	Exc, F	&	4abg	\\
2	-6	4	&	-6	2	-5	&	52	&	4	&	$	F_4	$	&	Exc, K	&	2aab	\\
-10	4	2	&	2	-5	-6	&	52	&	4	&	$	F_4	$	&	Exc, K	&	3aag	\\
3	-6	3	&	-2	1	-3	&	45	&	5	&	$	SO(10)	$	&	SO,T,F	&	2aab	\\
-9	3	3	&	1	-3	-2	&	45	&	5	&	$	SO(10)	$	&	SO,T,F	&	3aag	\\
3	3	-5	&	-2	-3	1	&	45	&	5	&	$	SO(10)	$	&	SO,T,F	&	5agb	\\
2	-4	2	&	-2	1	-2	&	28	&	4	&	$	SO(8)	$	&	SO,Exc	&	3aag	\\
2	2	-7	&	-2	-2	1	&	28	&	4	&	$	SO(8)	$	&	SO,Exc	&	4abg	\\
-6	4	3	&	2	-3	-4	&	21	&	3	&	$	SO(7)	$	&	SO,T,K	&	1aaa	\\
-3	3	4	&	-3	2	-4	&	21	&	3	&	$	SO(7)	$	&	SO,T,K	&	6baa	\\
-5	3	2	&	3	-5	-4	&	14	&	2	&	$	G_2	$	&	Exc, T	&	3aag	\\
-3	1	-3	&	-1	3	-1	&	3	&	1	&	$	3d_1	$	&	3d	&	1aaa	\\
-3	3	3	&	1	-1	-1	&	3	&	1	&	$	SU(2)	$	&	3d,T	&	1aaa	\\
-2	1	-4	&	-2	4	-1	&	3	&	1	&	$	SO(3)	$	&	3d	&	1aaa	\\
-3	-3	1	&	3	-1	-1	&	3	&	1	&	$	3d_1	$	&	3d	&	6baa	\\
-1	-5	1	&	5	-3	-1	&	3	&	1	&	$	3d_2	$	&	3d,T	&	6baa	\\
-1	3	3	&	-1	1	-1	&	3	&	1	&	$	SU(2)(1)	$	&	3d,T	&	6baa	\\
3	-1	1	&	-1	-1	1	&	3	&	1	&	$	SU(2)(2)	$	&	3d,T	&	6baa	\\
1	1	-3	&	1	-3	1	&	3	&	1	&	$	3d_1	$	&	3d	&	7bga	\\
3	-1	1	&	-1	1	-1	&	3	&	1	&	$	SU(2)	$	&	3d	&	7bga	\\
3	1	-1	&	1	-5	3	&	3	&	1	&	$	3d_2	$	&	3d,T	&	7bga	\\
-1	-1	-1	&	2	-1	-1	&	1	&	1	&	$	SO(2)	$	&	SO,$D_{2,1,\lambda}$	&	5agb	\\
\hline
	\end{tabular}
	\label{tab:is-phys}
\end{table}
\end{scriptsize}

Finally, let's consider 0/0 case. There are two different, up to permutations, cases: $2t-\alpha=0, \alpha=0$ and $2t-\alpha=0, \beta=0$. Assume we approach first point by line    $2t-\alpha=z\alpha$ as $\alpha \rightarrow 0$. Then $f(x) \rightarrow z$ at $\alpha \rightarrow 0$ and $dim  \rightarrow z$ at $\alpha \rightarrow 0$, since $dim=f(x)$ at $x \rightarrow 0$. Similarly, if we approach second point by line $2t-\alpha=z\beta$ as $\beta \rightarrow 0$, we have $f(x) \rightarrow z(q^2+1+q^{-2})$, $dim  \rightarrow 3z$  at $\beta \rightarrow 0$, where $q=e^{x\gamma/4}$. In both cases we get non-singular $f(x)$, but answer depends on how we approach a limiting point. An illustration is point (-2,4,0) which according to Table \ref{tab:par} corresponds to algebra SO(4), and according to formula for general SO(N) have a dimension N(N-1)/2=6. Actually this general N define a line, by which we approach the SO(4) point $2\alpha+\beta=0, \gamma=0$, that is the SO line $2\alpha+\beta=0$  which corresponds to $z=2$ above. From the other side, would we approach the same point by 3d line $2\alpha+\beta+\gamma=0$, which corresponds to $z=1$, we shall obtain dimension dim=3.

\section{Road map: roads.} \label{srmr}

In previous Section we study the population of Vogel plane, i.e. points corresponding to the interesting objects, those with non-singular universal character (\ref{gene2}). In the present Section we are interested in roads, connecting these populated points, by which we understand a straight lines, connecting sufficiently large numbers of them. The corresponding ``road's map'' turns to be rich, extending some features of corresponding map of simple Lie algebras. The main feature of latter is that algebras are all located along a few roads  intersecting at crossroads. Indeed, all SU(N) algebras are located along the line  $ \alpha+\beta=0$, which we shall call SU line, all SO(N) - along the line $SO: 2\alpha+\beta=0$, all Sp(2n) - along the line $Sp:\alpha+2\beta=0$. Moreover, taking into account the possibility to interchange parameters (which is another face of Sp(2n)=SO(-2n)), last two lines actually coincide, and we have two lines incorporating all classical groups. Finally, all five exceptional algebras lie on one line Exc: $ \gamma=2(\alpha+\beta)$.

The choice of lines is to a greate extent arbitrary, in a sense that we are free to put a line through any two points. But the line becomes worth to mention  when it contains a large number of populated points,  in analogy with classical series or exceptional algebras' lines. In our case we introduce the following lines:  T line with points satisfying $\alpha+2\beta=\gamma$, F line with equation $\alpha=\beta+\gamma$, K line $\alpha+2\beta=2\gamma$, M line $3\alpha=2\beta+2\gamma$ and (introduced earlier)D line $\alpha+\beta+\gamma=0$, 0d line $2\alpha+2\beta+\gamma=0$, 3d line $2\alpha+\beta+\gamma=0$.   In Tables \ref{tab:4abg}-\ref{tab:is-phys} for each point is shown whether it belongs to one of these lines.  We see that a lot of points belong to these lines, particularly many to the T line.This line $T$ is particularly interesting for us as a line containing the points $X_1$ and $X_2$, and it appears to coincide with the so-called subexceptional line in \cite{LM1}, denoted there $F3_3$.  

When restricted to particular line, universal expression for dimensions (\ref{f3}) has one parameter. Evidently, if it gives an integers at some values of that parameter, it will happen at some isolated points, which are easy to find, if one have the expression (\ref{f3}) and given line. Exactly in that way the point $E_{7\frac{1}{2}}$ was found in \cite{W,LM3}. It is not possible to do that without specifying a line, because simple requirement of being integer is unrestrictive for function of two variables (\ref{f3}) and will give as a solution the whole curves. 

One can consider the points of intersection of these lines. Many intersection points can be found from the Tables, namely in many cases one point belongs to more than one line, which means an intersection in that point. Actually tables completely answer on a question of intersections on a isolated points.  Note also that a  ``central'' point  $(\alpha,\beta,\gamma)=(1,1,1)$ (k,n,m)=(5,5,5) (in all patterns) doesn't belong to any of mentioned lines. Geometry of lines on Vogel's plane will be discussed elsewhere \cite{MV2}.

\section{$X_1,X_2$ and $E_{7\frac{1}{2}}$}   
\label{sxxe}

These three objects in the physical semiplane seems to be  interesting, first two of them were unknown. $E_{7\frac{1}{2}}$ was discovered in \cite{W,LM3} as a point on a exceptional line, where dimension formulae give integer answer. It is proved to be a Lie algebra, lying between $E_7$ and $E_8$ (hence notation) in a following sense: the six-dimensional algebra S  (``sextonions'') in between the quaternions and octonions is identified in \cite{W,LM3} (and earlier in \cite{Je,Kl}), and the triality construction of exceptional algebras is applied to S to get an algebra $E_{7\frac{1}{2}}$. It appears to be semidirect product $E_7\rtimes H_{56}$ where H is (56+1)-dimensional Heisenberg algebra, where \underline{56} is an irrep of $E_7$ with an invariant symplectic form. 

The dimension formulae  we checked  works for $X_1, X_2$ algebras, also. For example for $X_2$ we have $dimY_2(\alpha)=3927,dimY_2(\beta)=77,dimY_2(\gamma)=945$ , for $X_1$   $dimY_2(\alpha)=10166, dimY_2(\beta)=90,dimY_2(\gamma)=1989$. 
We have not identified, up to now, $X_1$ and $X_2$ as some Lie algebras. One can note however that their dimensionalities can be represented in a similar to $E_{7\frac{1}{2}}$ way: $dim(X_2)=99=66+32+1$, where 66 is the dimension of an algebra SO(12), 32 is dimension of  its spinor representation, and $dim(X_1)=156=91+64+1$, where 91 is dimensionality of algebra SO(14), 64 is  its spinor representation. Moreover, each of mentioned  spinor representations has an invariant symplectic form, in full analogy with \underline{56} of $E_{7\frac{1}{2}}$. In physics literature that form is called a charge conjugation matrix, see  \cite{KT,SS}. So, the natural hypothesis is that $X_1$ and $X_2$ are the semidirect products  $SO(14)\rtimes H_{64}$ and $SO(12)\rtimes H_{32}$ respectively, where $H_{64}, H_{32}$ are Heisenberg algebras based on an invariant symplectic forms in corresponding spinor representations. It is interesting that one of these algebras, $SO(12)\rtimes H_{32}$ (denoted there $D_6. H_{32}$) already appears in additional row/column of extended magic square in \cite{LM3,W}, although it was not known at that time that that entry corresponds to a point in Vogel's plane.

\section {Dimension formulae}  \label{sdf}
Since integers k, n, m and Diophantine conditions (\ref{tab:Diophantine}) are encoding identifying information about simple Lie algebras, it may be possible to express different quantities of that algebras through these integers. In Table \ref{tab:dim} we are presenting dimensions  formulae for all patterns.
\begin{small}
\begin{table}[ht]
\caption{Dimensions}
	\centering
		\begin{tabular}{|r|c|}   \hline
Pattern & dim \\ \hline
1aaa &	$	k - 3 m - 3 k m - 3 n - 3 k n + 4 m n$\\
2aab & $ n(k+1)(m-1)+k $\\
3aag &   $m(k-n-kn)  $ \\
4abg & $-knm$  \\
5agb & $-knm$\\
6baa & $k(3n+2m-nm)$ \\
7bga & $-knm$  \\  \hline
	\end{tabular}
	
	\label{tab:dim}
\end{table}
\end{small}

It is beautiful that it is possible to express dimensions as an explicitly integer numbers, without denominators of dimension formula (\ref{f3}). This gives some (partial) explanation, as promised above, why Diophantine conditions actually are sufficient to provide a regularity of character function f(x). 

We see from e.g. expressions for fourth, fifth and seventh patterns, that solutions of Diophantine equations directly give a multipliers in decomposition of dimension of algebra. For example, $dim(E_6)=78=-knm=-3\times 2 \times (-13)$, according to pattern 4abg, same number is represented in pattern 3aag in two ways - as $78=m(k-n-kn)=3\times 26$   and  $78=m(k-n-kn)=2\times 39$.

Dimensions of irreducible representations in a symmetric square of adjoint representations are given by \cite{V,Del,LM1}

\begin{equation}
 \label{Y}
 dim \, Y_2(\alpha)=-\left( \frac{\left( 3\,\alpha - 2\,t \right) \,\left( \beta - 2\,t \right) \,\left( \gamma - 2\,t \right) \,t\,\left( \beta + t \right) \,
      \left( \gamma + t \right) }{\alpha^2\,\left( \alpha - \beta \right) \,\beta\,\left( \alpha - \gamma \right) \,\gamma} \right)
 \end{equation}

and similar expressions, obtained by permutation of parameters. It would be interesting to present them in a similar form, i.e. have it expressed through k, n, m with denominators canceled.

\section{Conclusion} \label{sc}

We develop further the universal Lie algebra approach on the basis of previously obtained universal expression f(x) for character of adjoint representation, evaluated in a specific point $x\rho$. Requesting  a regularity of f(x) in finite complex plane as some general feature called to classify simple Lie algebras  within Universal Lie algebra approach, we get seven patterns of singularities cancellation. Each pattern gives a number of solutions for points on a Vogel plane, which include both simple Lie algebras and other points, with positive or negative dimensions. Most of simple Lie algebras appear in pattern 3aag, except SO(2N+1), which appears in patterns 2aab and 6baa. So, the minimal number of patterns needed to describe all simple Lie algebras is two. Patterns 3aag and 2aab also describe three  additional interesting points, the only additional ones appearing in physical semiplane - $E_{7\frac{1}{2}}, X_1, X_2$. For description of all other additional objects, belonging to unphysical semiplane and having negative dimensions, other patterns are needed, but in this case also some patterns can be discarded, e.g. 7agb. So, one can say that Vogel plane has seven sheets, which coincide in some points, namely those where the same objects are described in different patterns. For example all these sheets coincide at one point  $Y_1$ with parameters $(1,1,1)$, because it is present in all patterns. 
 
This  approach can be considered as an attempt to classify simple Lie algebras from the  universal Lie algebra side. It reveals an interesting connection between simple Lie algebras and certain Diophantine equations, those of Table \ref{tab:Diophantine}. 

The points in physical semiplane are almost all already known, except the $X_1, X_2$. We assume their similarity with previously known  $E_{7\frac{1}{2}}$ and make a hypothesis on what Lie algebras they can correspond to. The points $Y_i$ in unphysical semiplane cannot correspond to any Lie algebra due to negative dimensions. Two extremal points of view can be applied to these objects. One is that they are completely irrelevant and play no role. Other is that they can be considered on the same footing as those in a physical semiplane, since what we need finally from any Lie algebra are some numbers (e.g. eigenvalues of Casimirs) which describe some physical quantities. It is expectable that such a numbers can be calculated for $Y_i$ (which  amounts to substitute their values of parameters into universal expressions for that quantities, provided they exist).
It is not possible to make a final conclusion now, our point of view is that $Y_i$, as well as the whole universality approach worth further investigation.

For all the objects with regular f(x),  particularly for simple Lie algebras, expressions (\ref{gene}, \ref{gene2})  provide us with a universal interpretation for different roots of algebras, or more exactly their scalar products with Weyl vector $\rho$. This interpretation we consider as a step toward a universal expression for a group's volume, see \cite{MV}. It would be helpful for that purpose to have some algebraic formula for solutions of Diophantine equations above.

The Diophantine conditions, presented in Table \ref{tab:Diophantine} in both initial and normalised (i.e. without second-order terms) forms sometimes have a symmetry between some permutations of parameters k,n,m.

\begin{scriptsize}

\begin{table}[ht]
\caption{Diophantine equations}
	\centering
		\begin{tabular}{|r|r|c|r|} \hline
		Pattern&Diophantine Eq.:  kmn=&Transformation: (k,n,m)$\rightarrow$&Normalized form:  kmn=\\ \hline
		1aaa& mn+2kn+2km&(k,n,m)+(1,2,2)&4k+2m+2n+8\\
		2aab& mn+2kn+2km+2n-2k&(k,n,m)+(1,2,2)&2k+4n+2m+10\\
			3aag&mn+kn+2km+3n+2k&(k,n,m)+(1,2,1)&4k+4n+2m+12\\
			4abg&nm+km+kn+3m+3n+3k+5&(k,n,m)+(1,1,1)&4m+4n+4k+16\\
		5agb&mn+2kn+2km+2n+2m-3k-5&(k,n,m)+(1,2,2)&4m+4n+k+8\\
		6baa&2mn+2kn+2km-2n-3m&(k,n,m)+(2,2,2)&m+2n+4k+6\\
		7bga&2mn+2kn+2km-2n-2m-2k+5&(k,n,m)+(2,2,2)&2m+2n+2k+9\\  \hline
				\end{tabular}
	\label{tab:Diophantine}
\end{table}
\end{scriptsize}

For example, for the pattern 1aaa both forms are invariant w.r.t. the transposition $n \leftrightarrow m$. This transposition, acting on the expressions for $\alpha,\beta, \gamma$ interchange $\beta$ and $\gamma$, i.e. leads to the same object. But for the pattern 3aag situation is more complicated and interesting  \cite{MV2}. Namely, in normalised form we notice a symmetry of Diophantine condition under interchange of k and n. On the language of initial k and n, corresponding to initial form of Diophantine condition, it is a transformation $(k,n,m)\leftrightarrow(n-1,k+1,m)$. Under this transformation parameters $\alpha,\beta,\gamma$ really change their values, transposing descriptions of algebras given in Table \ref{tab:3aag}. For example, for points in physical semiplane we get transpositions $SU(2n)\leftrightarrow SO(2n+2), SU(k+1) \leftrightarrow SU(k),  X_1 \leftrightarrow E_8(1), F_4 \leftrightarrow E_6(2), SO(10) \leftrightarrow E_6(1), E_7 \leftrightarrow E_7, G_2\leftrightarrow SO(8), E_{7\frac{1}{2}} \leftrightarrow E_8(2), 0d_1 \leftrightarrow 0d_4, 0d_2 \leftrightarrow 0d_3, 0d_5 \leftrightarrow 0d_6$. This will be discussed elsewhere \cite{MV2}.

\section{Acknowledgments.}

I'm indebted to A.Veselov and H.Khudaverdian for encouraging discussions of present work. It is partially supported by the grant of Science Committee of Ministry of Science and Education of the Republic of Armenia.

\end{document}